\documentclass[10pt,letterpaper]{JHEP3}

\usepackage{amsmath}
\usepackage{amsfonts}
\usepackage{amssymb}

\usepackage{graphics}
\usepackage{amsmath}
\usepackage{amsfonts}
\usepackage{amssymb}
\usepackage{graphicx}
\usepackage{epsfig}

\def\ba{\begin{eqnarray}}
\def\ea{\end{eqnarray}}
\def\be{\begin{equation}}
\def\ee{\end{equation}}

\def\d{\mathrm{d}}

\def\mn{_{\mu \nu}}
\def\mupn{^\mu_{\, \nu}}
\def\({\left(}
\def\){\right)}
\def\eps{\epsilon}

\def\kf{\kappa_5^2}
\def\kq{\kappa_4^2}

\def\l{\ell}
\def\al{\left|\l\right|}

\def\by{\bar{y}}
\Roman{section}

\title{Mimicking $\Lambda$ with a spin-two ghost condensate}

\author{Claudia de Rham$^1$ and Andrew J. Tolley$^2$ \\
~$^1$ Ernest Rutherford Physics Building, McGill University, \\
~~~Montreal, QC H3A 2T8, Canada.\\
~~~E-mail: \email{\tt derham@hep.physics.mcgill.ca} \\
~$^2$ Joseph Henry Laboratories, Princeton University, Princeton NJ, 08544, USA.\\
~~~E-mail: \email{\tt atolley@princeton.edu}\\}

\abstract{We propose a simple higher-derivative braneworld gravity model which
contains a stable accelerating branch, in the absence of a cosmological constant or
potential,
that can be used to describe the late time cosmic
acceleration.
This model has similar qualitative features to that of
Dvali-Gabadadze-Porrati, such as the recovery of four-dimensional gravity at
subhorizon scales, but unlike that case, the graviton zero mode is
massless and
there are no linearized instabilities. The acceleration rather is
driven by bulk gravity in the form of a spin-two ghost condensate. 
We show that this model can be
consistent with cosmological bounds and tests of gravity.}

\begin{document}


\section{Introduction}

The cosmological constant problem remains one of the most challenging
problems in physics. This has
only been enhanced by the
recent observational data implying the presence of a late-time dark
energy component with $w=p/\rho$ tantalizingly close to $-1$ \cite{Riess:1998cb}. One line
of approach to making progress with this problem has been to
reinterpret the observed cosmic acceleration as being due to a
modification of gravity, thus setting the true cosmological constant
to zero. Although by itself this does not solve the problem, one may
then invoke the old argument that some symmetry or selection principle
may set $\Lambda$ to zero. For reviews on the cosmological constant
problems, see Refs. \cite{Weinberg:1987dv,Peebles:2002gy,Sahni:1999gb,Bludman:2006cg,Padmanabhan:2006ag}, and references therein.

One of the most interesting models of this kind was proposed by Dvali,
Gabadadze and Porrati (DGP) \cite{Dvali:2000rv,Dvali:2000km,Dvali:2000hr,Dvali:2000xg,Deffayet:2001pu} and has received
considerable attention in recent years (see Ref. \cite{Lue:2005ya} for a review). This model describes our
universe as a 3-brane localized in a five dimensional bulk where the
brane and bulk actions are taken to be pure Einstein-Hilbert (EH)
terms. As in many models of gravity, in which higher (or lower) scaling
dimension curvature terms are included, there can exist multiple branches of solutions. It was shown that one branch
corresponds to a self-accelerating one in which the brane geometry is
de Sitter (dS) which could describe the observed cosmic
acceleration.

The DGP model is unusual among higher dimensional models, in that
the graviton zero mode is massive. This implies that the graviton has
five rather than the usual two propagating degrees of freedom and has
significant implications for the physics of the model.
 For instance gravity can be shown to become nonlinear at intermediate
 scales
 \cite{Arkani-Hamed:2002sp,Rubakov:2003zb,Luty:2003vm,Nicolis:2004qq}. 
After some
debate in the literature it
has recently been demonstrated that the self-accelerating branch of the DGP model admits a ghost
state, see Refs.
\cite{Luty:2003vm,Nicolis:2004qq,Gorbunov:2005zk,Koyama:2005tx,Charmousis:2006pn}. This strongly suggests that
the
DGP model has both a classical and quantum instability.

Although a stable alternative to the DGP model has been suggested in
Ref. \cite{Padilla:2004tp}, it relies on the presence of an
asymmetric brane, for which the position of the brane represents a
new scalar degree of freedom.

In this work we propose a new model, similar in spirit to the DGP
model, which also admits a accelerating branch in the absence of a bulk or brane cosmological constant, but for which the
graviton zero mode is massless and no ghost states or unstable modes are present.
Our model also describes a 3-brane in a five dimensional bulk. The brane action contains an EH term and the bulk
contains an EH and a Gauss-Bonnet (GB) term $\mathcal{R}_2$.

The key feature of our model will be to take an unconventional form for the bulk action
\be
S=\frac{1}{2\kf}\int \d ^5 x
\sqrt{-g}\, \Big[-\mathrm{R}-\frac{\l^2}{2}\mathcal{R}_2\Big],
\ee
where the EH term has the wrong sign. If we expand around the
canonical branch vacuum, {\it i.e.} Minkowski space, the graviton kinetic
term has the wrong sign and will lead to instabilities in the presence
of bulk matter. By contrast if we expand around the non-canonical Anti-de Sitter
(AdS) vacuum branch, the graviton kinetic term is conventional and
no instability occurs at the linearized level. It is this unconventional
branch that we shall consider in this work.

This model may be understood as a type of spin-two ghost condensation \cite{Arkani-Hamed:2003uy}.
Since the bulk EH term has the wrong sign, bulk gravitons act like
ghosts, but by expanding around a non-trivial vacuum we recover a
stable theory. In the ghost condensation work, the non-trivial
solution was taken to be a time-dependent solution, thus breaking
Lorentz invariance \cite{Cheng:2006us}. In the present case, de
Sitter invariance on the brane is maintained by taking a solution
which depends on the bulk coordinate $y$.

In what follows we will:
\begin{enumerate}
\item Present a summary of why the self-accelerating branch of DGP
  contains a ghost instability, and relate it with the different
  limits of the Randall-Sundrum (RS) model with an additional EH term on the brane.
  In particular we will show that the unstable
  DGP branch is related to a RS brane embedded in an
  infinite extra-dimension with warp factor diverging away from the
  brane, for which gravity is not localized.
\item We will then propose a new model which admits an accelerating
  branch, which we show is {\bf stable at the linearized level}. We emphasize that no cosmological
  constant is introduced on either the bulk or the brane.  The
  graviton zero mode is shown to be massless and the `dangerous' brane-bending modes decouple.
\item At subhorizon scales, we show that the model effectively gives rise
  to four dimensional gravity at the linearized level.
\item To finish, we fine-tune the model in order to obtain the correct
  order of magnitude for the current cosmic acceleration on the
  brane. Ensuring that we get conventional cosmological evolution from
  nucleosynthesis onwards, requires
  that the five dimensional Planck mass is remarkably small ($\approx
  10$ eV). We argue that nevertheless,
  this can be acceptable since gravitation is dominated by the brane
  EH  term, and five-dimensional effects such as small black hole production are
  suppressed.
\end{enumerate}

\section{Stability of the RS$_{\pm}$ model}

In this section we study the
stability of the modified RSII model \cite{Randall:1999vf}, to prepare
the ground for the new model presented in section \ref{newmodel} and
to gain insight into the DGP model. Readers more familiar with these
techniques can skip straight to section \ref{newmodel}.
In the RS$_+$ (resp. RS$_-$) model, the extra-dimension has a warp
factor that falls off (resp. diverge) exponentially away from the
brane.
In both models, we include a EH term on the
brane:
\ba
S=\frac{1}{2\kf}\int \d ^5 x
\sqrt{-g}\Big[\mathrm{R}-2\Lambda\Big]+ \int \d^4x
\sqrt{-q}\left[\frac{\alpha}{2\kq}R+\mathcal{L}_m-\lambda+
\frac{1}{\kf}K\right],
\ea
where $\mathrm{R}$ is the five-dimensional scalar curvature while $R$
is the four-dimensional one induced on the brane. The bulk has a
negative cosmological constant $\Lambda$. $\mathcal{L}_m$ is the
Lagrangian for the matter fields confined to the brane and $K$ is the
trace of the extrinsic curvature. The
parameter $\alpha$ is to be associated to the cross-over scale $r_0$ introduced in the DGP
model,
$\alpha=r_0\kq/\kf$. In what follows, we consider $\alpha>0$, unless
specified differently.

When the bulk has a negative AdS curvature (RS$_-$ model), we will show that the
geometry on the brane is self-accelerating,
but gravity is not localized on the brane, whereas when the AdS
curvature is positive (RS$_+$), the graviton zero mode is massless but
 gravity is not self-accelerating. We then
discuss the consequences of these models when the curvature vanishes
($\Lambda\rightarrow 0$), and draw a parallel with the two branches of
the DGP solution.

\subsection{Background behavior and self-accelerating solution}

In the presence of a negative cosmological constant, the bulk geometry
is AdS with curvature scale $|\l|$ set by $\Lambda=-6/\l^2$.
Since we are interested in solutions in which the brane geometry is
accelerating,
we choose the dS slicing of AdS
\be
\d s^2=a^2(y)\(\d y^2+\gamma_{\mu\nu}\d x^{\mu}\d x^{\nu} \)=
\frac{\l^2H^2}{\sinh^2 Hy} \(\d y^2-\d t^2+e^{2Ht} \d\vec{x}\, ^2
\), \label{AdS dSslicing}
\ee
where the geometry is compactified on a $S^1/\mathbb{Z}_2$-orbifold,
with a brane located at the fixed point $\by$ of the symmetry. $\by$
is such that $\sinh H\bar{y}=\l H$, with $H$ the Hubble constant on the brane. The induced
metric on the brane is hence $\gamma\mn$.
Curved braneworlds have been considered in Refs. \cite{Kaloper:1999sm,Bowcock:2000cq}.

For convenience we shall allow $\l$ to take both signs
and consider only the region $y \ge \bar{y}$ of the $\mathbb{Z}_2$
orbifold (the metric in its copy being obtained by $y \rightarrow 2\bar{y}-y$).
If $\bar{y}>0$, the region is then bounded by the horizon at $y=+\infty$. If $\bar{y}<0$
then the region is bounded by the boundary of AdS at $y=0$. Denoting
$\l=-\epsilon |\l|$, by taking $\epsilon=-1$ we describe the RS$_+$ model while taking
$\epsilon=+1$ represents the RS$_-$ one.

The extrinsic curvature is given by
$K\mupn=-\frac{1}{\l}\sqrt{1+\l^2H^2}\, \delta\mupn$ and is
determined by the Isra\"el junction conditions \cite{Israel:1966rt} to be
\be
K\mupn=-\frac{\kf}{6} \lambda
\delta\mupn-\frac{\kf}{2}\(
T\mupn-\frac{1}{3}T\delta\mupn \)+\frac{\alpha\kf}{2\kq}
\left( R\mupn-\frac{1}{6}R\delta\mupn \right),
\ee
where $T\mn$ is the stress-energy tensor associated with $\mathcal{L}_m$.
We take the canonical value for the brane tension: $\lambda=6/\kf
\l$, and hence in our convention, the RS$_+$ model will have a
positive tension $\lambda>0$ brane while RS$_-$ has $\lambda<0$. We
emphasize that due to the presence of EH term on the brane, this
choice is arbitrary and we could have also have chosen both
RS$_\pm$ models to have a vanishing tension brane, or a tension with
opposite sign. This is only a matter of convention. In section
\ref{newmodel}, we shall be interested in a model where the brane
tension vanishes, and this convention will therefore play no r\^ole.

 When some matter fields with energy density
$\rho=-T^0_{\, 0}$ are confined to the brane,
the junction
conditions imply
\be
-\frac{1}{\l}\sqrt{1+\l^2H^2}=-\frac{\kf}{6}\(\lambda+\rho\)+\frac{\alpha
  \kf}{2\kq}H^2,
\ee
that is,
\be
\sqrt{1+\l^2H^2}=1-\eps \frac{\kf\al}{6}\rho
+\frac{\eps}{3}\gamma\, \l^2 H^2 , \label{Fried1}
\ee
where we write $\gamma=\frac{3\alpha\kf}{2\al \kq}>0$.
We may note that in this formalism, \eqref{AdS dSslicing} is a solution of the equations of motion
only if the Hubble parameter is constant, {\it i.e.} only if
$\rho$ is a cosmological constant. However a more detailed calculation
shows that the final Friedmann equation \eqref{Fried1} remains valid
for any kind of matter (see for instance
Refs. \cite{Bowcock:2000cq,Padilla:2002tg} for details of this
derivation for arbitrary kinds of matter).

\begin{itemize}
\item For the branch $\eps=-1$, this equation admits only one solution for real
$H^2$. At low-energy, this solution is similar to the
usual Friedmann equation:
$H^2=\kq \rho /\al \(3+2\gamma\)$,
and when no matter in present on the brane, $H=0$. This solution is
therefore the conventional one and is not self-accelerating.
\item For $\epsilon=+1$, this equation admits two branches of solutions, the
canonical one for which $H=0$ when $\rho=0$, and a non-canonical one for which
\be
H=\frac{2 \kq}{\alpha \kf} \, \sqrt{1-\frac{2}{3}\gamma},\label{H -}
\ee
when $\gamma<3/2$.
However, if
$\gamma$
satisfies this inequality, the canonical solution for which $H=0$ when $\rho=0$
couples the wrong way to matter: At low-energy
$H^2=-\kf \rho / \al \(2 \gamma-3\)$,
and is not an acceptable
solution.  Thus there is only one physical branch for each sign of
$\epsilon$.
\end{itemize}

\subsection{Perturbations}

\label{tensors1}

To determine the stability of the model, we first consider the metric
perturbations in the RS gauge
\be
\d s^2=a^2(y)\left( \d y^2+\gamma_{\mu\nu}\d x^{\mu}\d x^{\nu}
\right)+\sqrt{a}\,  h_{\mu\nu}(y,x^{\mu})\,  \d x^{\mu}\d x^{\nu}. \label{ds pert}
\ee
As is usual for perturbations around a maximally symmetric spacetime,
there is sufficient remaining  gauge freedom to set $ h_{\mu\nu} $ to
be transverse and traceless with respect to the
dS metric $\gamma_{\mu\nu}$. This is the statement that there are no propagating
vector and scalar modes in the bulk, in the absence of bulk matter.
In what follows we raise indices as $h\mupn=\gamma^{\mu \alpha}h_{\alpha\nu}$.
In this gauge, the perturbed equations of motion in the absence of any matter in
the bulk are
\ba
\left[ \partial_y^2 +\boxdot
  -\left(\frac{9}{4}+\frac{15}{4\sinh^2 H y} \right)H^2 \right]
h^{\mu}_{\nu}=0,\label{eq h bulk}
\ea
with the notation $\boxdot=\left[\Box-2H^2\right]=\left[\gamma^{\mu\nu}\nabla_\mu
\nabla_\nu-2H^2\right]$, where $\nabla_\mu$ is the covariant derivative with respect to
 $\gamma\mn$.
Note that in this
gauge, the brane is not static but located at $y=\bar{y}+\delta y(x^\mu)$.
In order to derive the boundary conditions, we
work instead in the Gaussian Normal (GN) frame where the brane is
static. Performing the coordinate transformation,
\ba
y &\rightarrow& \tilde{y}=y-\frac{\delta y}{a(y)}, \\
x^\mu &\rightarrow&
\tilde{x}^\mu=x^\mu+\frac{\cosh H y}{\l H^2}\gamma^{\mu
  \nu}\delta y_{, \nu},
\ea
the perturbed metric in GN gauge is then
 \be
 \tilde{h}_{\mu\nu}=h_{\mu\nu}+2\epsilon \sqrt{\frac{|\l|}{H}}\frac{\cosh H y}{(\sinh
 H |y|)^{3/2}} \, \hat{F}\mn\delta y, \label{h GN y}
 \ee
 where $\hat{F}\mn=\left[\nabla_\mu \nabla_\nu+H^2
 \gamma\mn\right]$, and the metric perturbation induced on the brane
 is therefore:
\ba
\tilde{h}_{\mu\nu}(\by)=h_{\mu\nu}(\by)-\frac 2 H \coth H \by \ \hat{F}\mn\delta
y.
\ea
 In the vacuum, the perturbed boundary condition are then
 \ba
 \delta K\mupn = \frac 1 2 \left[\partial_y+\frac{3H}{2}\coth H \by
 \right]\tilde{h}\mupn(\by) = -\frac{\alpha \kf}{4 \kq}\boxdot
 \tilde{h}\mupn(\by). \label{bdyGN}
 \ea
Translating back to RS gauge, the boundary constraint is therefore
\ba
\left[\partial_y
+\frac{3H}{2}\coth H \bar{y}
+\frac{\alpha \kf}{2\kq} \boxdot \right]
h^{\mu}_{\nu}(\bar{y})=\(2
+\frac{\alpha \kf}{\kq H} \coth H\bar{y} \boxdot
\) \hat{F}\mupn\, \delta y . \label{bdyRS}
\ea
Since $h\mn$ is traceless, the
position of the brane satisfies: $\hat{F}^\mu_{\, \mu}\, \delta
y=\left[\Box +4H^2\right]\, \delta y=0$ in the vacuum, or in other words,
\ba
\boxdot\  \hat{F}\mn \delta y=2 H^2  \hat{F}\mn \delta y. \label{eq d y}
\ea
 Both the bulk equation \eqref{eq h bulk} and
the boundary equation \eqref{bdyRS} are identical for both branches $\eps=\pm 1$,
but we recall, that the only difference arises in the sign of
$\bar{y}$, and in the normalization of the mode functions. When
$\eps=-1$, the mode functions should be normalized over the region
$\bar{y}<y<\infty$, which requires them to fall as $y\rightarrow
\infty$ or behave as plane waves, whereas for $\eps=+1$, the mode functions should be
normalized over the region $\bar{y}<y<0$, which requires them to be finite as
$y\rightarrow 0^{-}$.
\begin{itemize}
\item Tensor perturbations
\end{itemize}
We can discard for now
the contribution from the brane displacement and concentrate on the
tensor perturbations.
Following the approach of Ref. \cite{Charmousis:2006pn} we write
$h_{\mu\nu}=\sum_m u_m(y) \chi_{\mu\nu}^{(m)}$ where
$\chi_{\mu\nu}^{(m)}$ are a complete set of transverse traceless
tensors satisfying
\be
\boxdot \chi_{\mu\nu}^{(m)}=m^2\chi_{\mu\nu}^{(m)}.
\ee
Then the mode functions $u_m(y)$ satisfy
\ba
&&\left[ \partial_y^2 +m^2 -\left(\frac{9}{4}+\frac{15}{4\sinh^2 Hy}
  \right)H^2 \right] u_m(y)=0, \label{bulkpm}\\
&&\left[\partial_y+\frac{3H}{2} \coth H\bar{y}+\frac{\alpha \kf}{2\kq}
  m^2 \right] u_m(\bar{y})=0.\label{bdypm}
\ea
It is straightforward to show that the background solution is a consistent solution of the
equation of motion and boundary condition if $m=0$:
\be
u_0(y)=A(|\sinh Hy|)^{-3/2}. \label{u0}
\ee
For $\epsilon=-1$, this mode is clearly
normalizable in the region $\bar{y}<y<\infty$. Consequently it describes the massless graviton zero
mode. It can be shown that no other normalizable modes satisfy the
boundary condition  \eqref{bdypm}, for $0<m^2 \le 9H^2/4$.
The remaining modes are therefore massive and satisfy $m^2 >
9H^2/4$. They represent the continuum of Kaluza-Klein modes \cite{Garriga:1999yh}.

For $\epsilon=+1$, {\it i.e.} when $\bar{y}<0$, the massless mode \eqref{u0}
is no longer normalizable
in the range $0>y\ge \bar{y}$ since it diverges at $y=0$.
In fact all the modes split into normalizable modes behaving as $(-y)^{5/2}$
as $y \rightarrow 0$ and non-normalizable modes behaving as  $(-y)^{-3/2}$.
This of course is no surprise since this is the familiar behavior of modes near the boundary of AdS.
Demanding that the modes are normalizable effectively quantizes the
masses $m$ and projects out the non-normalizable massless
mode. In appendix \ref{app neg}, we show that when no matter is
present on the brane, the zero mode has a mass $m^2>2H^2$. Although
this zero mode does not sit in the forbidden region $0<m^2<2H^2$
\cite{Higuchi:1986py,Higuchi:1989gz,Bengtsson:1994vn,Kogan:2000uy,Deser:2001pe,Deser:2001wx}
when
the brane is empty, it will if sufficient matter is introduced on the
brane. In other words, the brane in the RS$_-$ model becomes
unstable once the energy density confined on it is too
important. Furthermore, since the graviton is
not massless in this theory, gravity will not be
localized on such a brane. In particular, we show in appendix
\ref{app neg} that at subhorizon scales, four-dimensional gravity is
not recovered on the brane.

\begin{itemize}
\item Brane-bending mode
\end{itemize}
Let us now consider the mode which is sourced by the brane-bending
term $\delta y$. In order to understand this mode we follow the
arguments of Refs. \cite{Koyama:2005tx,Charmousis:2006pn}.
The equation of motion \eqref{eq d y} of the brane-bending mode can
be associated to the one of a transverse traceless tensor with mass $m^2=2H^2$. Thus
although the brane-bending mode is in a sense a scalar mode,
it is here degenerate with the tensor sector.

Solving the bulk equation for $m^2=2H^2$ we find the general
solution
\ba
h^{(2)}_{\mu\nu}= \(A \frac{\cosh
  Hy}{(\sinh Hy)^{3/2}}+B \frac{\(\cosh Hy-1\)^2}{(\sinh Hy)^{3/2}}\)\chi^{(2)}_{\mu\nu}.
\ea
The first solution proportional to $A$ behaves identically to the
brane-bending mode, whilst the $B$ mode is distinct.

For the RS$_+$ model, the B mode is non-normalizable
due to the divergence as $y \rightarrow \infty$. Consequently any
normalizable solution would have to be composed of the A mode
only. This combined together with the brane-bending mode gives the
metric perturbation in the GN gauge
\be
\tilde{h}^{ (2)}_{\mu\nu}=\( A \chi^{(2)}_{\mu\nu}
-2\sqrt{\frac{\l}{H}} \hat{F}\mn\delta y
 \) \frac{\cosh Hy}{(\sinh Hy)^{3/2}}.
\ee
However the boundary condition \eqref{bdyGN} acting on this mode simply sets the
combination $\(A \chi^{(2)}_{\mu\nu} -2\sqrt{\l H}\hat{F}\mn \delta y\)$
to zero.
Thus the brane-bending decouples for the tensor sector and
plays no further r\^ole.

By contrast, the RS$_-$ model ($\eps=+1$) requires
normalizability at $y \rightarrow 0$ which picks out the solution
\be
h^{(2)}_{\mu\nu}=B \, \chi^{(2)}_{\mu\nu} \frac{(\cosh Hy-1)^2}{(\sinh
Hy)^{3/2}}.
\ee
Then applying the boundary conditions, this mode couples together with
the brane-bending mode to give a nontrivial solution
\be
\tilde{h}^{(2)}_{\mu\nu}=\frac{2}{H} \(Z \, \frac{(\cosh
  Hy-1)^2}{(\sinh Hy)^{3/2}}- \sqrt{\l H} \frac{\cosh Hy}{(\sinh Hy)^{3/2}}\) \hat{F}_{\mu\nu} \delta y,
\ee
where
\ba
Z=\frac{1+\frac{\alpha \kf}{\kq}H \coth H \bar{y}}{\frac{2
    }{\(H\l\)^{5/2}}\(4+\cosh H \bar y\)\sinh^4\! \frac{H \bar
    y}{2}+\frac 3 2  \coth H \bar y+\frac{\alpha \kf}{\kq}H}.
\ea
Thus the brane-bending mode represents a genuine physical degree
of freedom. This is exactly like the DGP scenario where the
significance of this mode is explained in more detail in the
literature, (see Refs. \cite{Koyama:2005tx,Charmousis:2006pn} for the most recent and detailed
studies). 
When the matter content on the brane is such that the mass of
the graviton is $0<m^2<2H^2$, the zero helicity mode of the graviton
$\chi\mn^{(\text{h}=0)}$
is a ghost. In the case where $m^2>2H^2$, $\chi\mn^{(\text{h}=0)}$
recovers the conventional kinetic sign (this mode is
stable), but the brane-bending mode becomes a ghost. For the critical
mass of the zero mode $m^2=2H^2$, the coefficient in front of the
kinetic term of $\chi\mn^{(\text{h}=0)}$ vanishes and
the brane-bending mode takes the r\^ole of the helicity-zero mode.
In this case the brane-bending mode is a ghost. As long as the zero
mode is massive, a ghost is thus present in the theory.

\subsection{Flat bulk limit and analogy with DGP}

Let us now consider the limit $|\l| \rightarrow \infty$,
($\gamma\rightarrow 0$), so that both
the brane tension and bulk cosmological constant vanish which is the
limit in which we get the DGP model.
The stable branch of the DGP
model corresponds to the $|\l| \rightarrow \infty$ limit of the
$\epsilon=-1$ RS model. It retains the same features, namely that the
graviton zero mode is massless.
The unstable self-accelerating branch corresponds to the $|\l|
\rightarrow \infty$ limit of the $\epsilon=+1$ model.  We find
\be
H=\frac{2\kq }{\alpha \kf}.
\ee
The fact that we do not get a stable theory and a massless graviton
can be seen as a consequence of the fact that a brane in an
uncompactified AdS bulk which diverges away from the brane is pathological.

In the limit $|\l|
\rightarrow \infty$ in which case $\bar{y} \rightarrow -\infty$,
the equations of motion effectively become:
\ba
&&\left[ \partial_y^2 +m^2 -\frac{9}{4}H^2 \right] u_m(y)=0, \\
&&\left[\partial_y-\frac{3H}{2} +\frac{\alpha \kf}{2\kq} m^2 \right] u_m(\bar{y})=0.
\ea
This is the DGP limit and, as shown in Refs.
\cite{Koyama:2005tx,Charmousis:2006pn},
the mass of the normalizable zero mode is
$m^2=2\frac{\kq}{\alpha \kf}(3H-2\frac{\kq}{\alpha \kf})$. When no
matter is present on the brane, $H=2\kq/\alpha\kf$, and the mode has
the critical mass $m^2=2 H^2$. But as soon as some matter is present
on the brane, $H>2\kq/\alpha\kf$, and the modes lies then in the
forbidden mass range $0<m^2<2H^2$ which gives rise to ghosts in the scalar sector.

Although the two branches of the DGP model are two natural
solutions of the same model, they can be seen as the limit of two
very distinct models \cite{Deffayet:2000uy}. The stable branch of DGP $\eps=-1$, is the
limit when $\al\rightarrow \infty$ of the RS$_+$ model, which is a
well-defined stable model. The self-accelerating branch
$\eps=+1$ of DGP, is on the other hand the limit when
$\al\rightarrow
\infty$ of the RS$_-$ model.  Unlike in the RS$_+$ case, the zero mode
is not normalizable and the brane-bending mode does not decouple. This
is the key element, responsible for the instability
of the RS$_-$ model, and by continuity, to the instability of the $\eps=+1$ branch of
the DGP scenario.

As a short remark, we briefly
discuss here the situation when $\alpha<0$. In that case, we see that
the self-accelerating branch is associated to the RS$_+$ model
$\eps=-1$, where the size of the extra-dimension is finite:
$H=H_0=\frac{2\kq}{\left|\alpha\right|\kf}$.
In this case we therefore expect the normalizable zero mode to be massless, and hence
to obtain a stable self-accelerating solution. However this model is
pathological, since matter couples the wrong way to gravity, and this
solution will therefore be unstable against matter on
the brane. In other words there are ghosts in the brane matter sector. At low-energies, we have
$H^2=H_0^2-\frac{2\kq}{3\left|\alpha\right|} \rho$.

\section{New Model}

\label{newmodel}

In the previous sections we have seen that the presence of ghosts is
intimately connected with the infinite volume of the extra-dimension,
{\it i.e.} with the fact that the warp factor blows up away
from
the
brane.
In
this section we propose a new model in which the extra-dimension has
finite volume, and the warp factor falls off at infinity, but the brane geometry will be forced to accelerate
in the absence of a cosmological constant. The key is to modify the
bulk action and replace it with the action for a spin-two ghost
condensate. To understand this, let us first consider a similar
scalar field system defined by the action
\be
S=\int \d^5x \sqrt{-g} \( +\frac{1}{2} (\nabla \phi)^2 - \frac{\theta}{4} (\nabla
\phi)^4 \).
\ee
The conventional vacuum is $\phi=\phi_0=\text{const}$ in which case
the fluctuations $\delta \phi$ are ghosts.
However, if we consider the scalar field to live in five-dimensional
Minkowski space and perturb around the background solution
$\phi_b=\sigma\ y/\sqrt{\theta} $ so that $\phi=\phi_b+\delta \phi$,
the perturbed 
action is then
\be
\delta_2 S=\int \d^5x \sqrt{-g}\left[- \frac{1}{2}\Bigg(
(\sigma^2-1)\partial_{\mu}\delta \phi\, \partial^{\mu}\delta
\phi+(3\sigma^2-1)(\partial_y \delta \phi)^2\Bigg)\right].
\ee
It is clear that provided $|\sigma|>1$ the kinetic term will have
the conventional sign and the perturbations will be stable. The
drawback of this example is that five-dimensional Poincare invariance is broken
down to four-dimensional. By contrast, in the gravitational version discussed
below we will be able to perturb around an AdS background {\it
without} breaking the $SO(4,2)$ AdS isometry group.

\subsection{Accelerating solution}
\label{selfGB}

The gravitational version of the ghost condensate is defined by the
action
\ba
S=\frac{1}{2\kf}\int \d ^5 x
\sqrt{-g}\Big[-\mathrm{R}-\frac{\l^2}{2}\mathcal{R}_2\Big]+ \int
\d^4x \sqrt{-q}\left[\frac{\alpha}{2\kq}R+\mathcal{L}_m-\frac{1}{
  \kf}Q\right],\label{action}
\ea
where $\l$ is an arbitrary length scale, and the dimensionless
constant $\alpha$ is positive. The unconventional sign for the EH
term is balanced out by the larger GB term so that perturbations
around the AdS background solution (with length scale $\l$) have a
positive kinetic term. We stress that the GB term plus its boundary
term only introduces terms in the action of the form $(\partial_\mu
g_{\nu\omega})^4$ and so the equations of motion remain second
order. In particular this means that there are only five physical
fluctuations representing the five components of a graviton.

In section
\ref{implications}, we will consider the limit where the five
dimensional effects are highly suppressed
$\kf\gg \l \kq$ (or in other words, when the brane is very heavy). In
that limit, we need $|\alpha-1|\ll 1$ in order to recover the
right coupling of matter to gravity,
but for now we leave the three free parameters
$\l, \kf$ and $\alpha$ arbitrary.
The GB term $\mathcal{R}_{2}$
is the trace of
\ba
\hspace{-10pt}\mathcal{R}^{\,A}_{2\, B}=\mathrm{R} \, \mathrm{R}^{\!A}_{
 B}-2
 \,\mathrm{R}^A_ C\,\mathrm{R}^C_B-2  \mathrm{R}^{CD}\,\mathrm{R}^A_{\ \
   CBD} +\mathrm{R}^A_{\ \ DEF}\, \mathrm{R}_B^{\ \ DEF} ,\notag
\ea
$\mathrm{R}^A_{\ \ BCD}$ being the five-dimensional Riemann tensor.
The boundary term $Q$ is the generalization of the
Gibbons-Hawking boundary terms \cite{Gibbons:1976ue,Myers:1987yn}:
\ba
Q=K+ \l^2 \(J-2 G^\mu_\nu K^\nu_\mu\),
\ea
where $K\mupn$ is the extrinsic curvature on the brane, $G\mupn$
being the Einstein tensor on the brane and
\ba
J\mupn=-\frac 2 3 K^\mu_\alpha K^\alpha_\beta K^\beta_\nu+\frac 2 3 K\,
K^\mu_\alpha K^\alpha_\nu+\frac 1 3 K\mupn (K^\alpha_\beta K^\beta_\alpha-K^2).
\ea
In the five-dimensional bulk, the Einstein equations are
\ba
\mathcal{G}^A_B=-\mathrm{G}^A_B- \l^2\(
\mathcal{R}^{\,A}_{2\, B}-\frac{1}{4}
\mathcal{R}_{2}\,\delta^A_B\)=0.\label{Ein5d}
\ea
As already pointed out in
\cite{Deser:1989jm,Deser:1987uk},
at the background level, this modified Einstein equation has an
AdS solution even though no cosmological constant is introduced in
the bulk.
In the dS slicing, the metric is expressed in
\eqref{AdS dSslicing}.
Although $\l$ could a priori be negative, we concentrate in the rest of
this paper to the situation where $\l$ is positive: $\l=-\eps |\l|$,
with $\eps=-1$, using the same notation as the previous section.

The boundary conditions on the brane is given by the analogue of the
Isra\"el matching conditions \cite{Israel:1966rt} in presence of GB terms
\cite{Myers:1987yn,Neupane:2001kd,Davis:2002gn}:
\ba
&&\hspace{-50pt}-K^\mu_\nu-\frac{2 \l^2}{3}\(\frac 9 2 J^\mu_\nu- J
\delta^\mu_\nu-3P^\mu_{\ \ \alpha \nu \beta}K^{\alpha
  \beta}+ P^\rho_{\ \ \alpha \rho \beta}K^{\alpha
  \beta}\delta\mupn \)\notag \\
&&\hspace{120pt}=-\frac{\kf}{2}\(T\mupn-\frac 1 3 T \delta
\mupn\)+\frac{\alpha}{2}\frac{\kf}{\kq}\(R\mupn-\frac 1 6 R \delta \mupn\),
\label{bdyCond}
\ea
where
\ba
P^\mu_{\ \ \alpha \nu \beta}=
R^\mu_{\ \ \alpha \nu  \beta}+
\(R^\mu_\beta \, q_{\alpha \nu} +
R_{\alpha\nu}\, \delta^\mu_\beta-R_{\alpha
  \beta}\, \delta^\mu_\nu-R\mupn \, q_{\alpha \beta}\)
-\frac 1 2 R\(\delta^\mu_\beta\,
q_{\alpha\nu}-q_{\alpha \beta}\, \delta\mupn\),
\ea
$R^\mu_{\ \ \alpha \nu \beta}$ being the four-dimensional Riemann
tensor induced on the brane.
For cosmological implications on ordinary
five-dimensional EH-GB models, see Refs. \cite{Nojiri:2000gv}.

Working with the AdS bulk, the boundary condition \eqref{bdyCond} therefore reads:
\ba
\frac{1}{\l}\sqrt{1+\l^2H^2}\(1+4\l^2H^2\)=-\frac{\kf}{2}\rho+\frac{3\alpha}{2}\frac{\kf}{\kq}
H^2.\label{FriedGB}
\ea
Provided $\gamma=\frac{3\alpha}{2}\frac{\kf}{\l
  \kq}>\gamma_c$, (with the critical value
  $\gamma_c=\sqrt{\frac{207}{8}+\frac{33\sqrt{33}}{8}}\simeq7.04$),
  the
brane geometry is accelerating in the absence of any stress-energy or
cosmological constant on the brane.
Unlike the previous model for which $\gamma$ was bounded $\gamma<3/2$,
here $\gamma$ can be as large as we want.
When the brane is empty, the
  geometry is indeed dS, driven by an effective cosmological constant
\ba
\Lambda_{\text{eff}}=\frac{3\bar{r}}{\l^2\kq}, \label{cc2}
\ea
where $\bar{r}$ is the solution of  $\gamma r
  =\sqrt{1+r}\(1+4r\)$. For  $\gamma>\gamma_c$, this equation has two
  real positive solutions, we take $\bar{r}$ to be the smallest
  one. For $\gamma \gg 1$, this is
\ba
\Lambda_{\text{eff}}\sim \frac{2}{\alpha \kf \l}. \label{cc}
\ea
To understand the coupling to matter, we use the notation
\ba
2\l \kf\ \rho_\text{higher}&=&\sqrt{1+\l^2H^2}\(1+4\l^2H^2\)-\(1+\frac{9}{2}\l^2H^2\)=\mathcal{O}(\l^4
H^4).
\ea
The Friedmann equation then takes the form
\ba
H^2=\frac{1}{\l^2\(\gamma-\frac 9 2 \)}+\frac{\kf}{2\l\(\gamma-\frac
   9 2\)}\, \(\rho+\rho_{\text{higher}}\),
   \label{H2}
\ea
where the higher order terms are negligible
if $\l H\ll 1$.
In this limit, we recover the conventional coupling to matter in the
Friedmann equation, if we impose the relation
$\kf/2\l\(\gamma-\frac
  9 2\)=\kq/3$, {\it i.e.}  if
\ba
\alpha=1+\frac{3\l\kq}{\kf}=
  1+\frac{9}{2\gamma}+\mathcal{O}(\gamma^{-2}).
\ea
When $\l H\ll1$, (or equivalently, at low-energy, or when $\gamma\gg1$), we then have
\ba
H^2=\frac{\kq}{3}\,\(  \rho +\Lambda_{\text{eff}}\)+O(\rho^2). \label{H3}
\ea
We discuss the viability of this cosmological evolution in section
\ref{implications}. We may note again, that although this Friedmann
equation was derived using the assumption that $H$ was constant, (and
hence for constant $\rho$), this result remains valid beyond this
assumption and is generic for any kind of matter on the brane (see for
instance Ref. \cite{Charmousis:2002rc}).

\subsection{Perturbations and Stability}
Let us now consider perturbations around the bulk AdS solution in
the RS gauge \eqref{ds pert}.
This follows closely the discussion in section
\ref{tensors1}.
In order to understand, how the bulk equation is affected by the
presence of GB terms and by the overall negative sign in the bulk
action, let us first consider the case where the bulk action is
\ba
S^{(\xi, \beta)}=\int \d^5 x\sqrt{-g}\left[\frac{\xi}{2 \kf}\(
  \mathrm{R}+\frac{\beta \l^2}{4}\mathcal{R}_2\)+\mathcal{L}_{\text{bulk}}\right],
\ea
the Einstein equation gets an overall $\xi$ factor.
The GB terms give an additional contribution corresponding to an extra $(1-\beta)$ factor,
(see Refs. \cite{Deser:1989jm,Neupane:2001st} or  Eq.(17) of
Ref.\cite{deRham:2006hs}). The resulting Einstein equation is therefore
\be
\xi\(1-\beta\)\left[ \partial_y^2 +\boxdot -\left(\frac{9}{4}+\frac{15}{4\sinh^2
    Hy} \right)H^2 \right] h\mupn=-\frac{\kf}{2}\tau \mupn.
\ee
where here again $h\mupn$ is transverse and traceless with respect to
$\gamma\mn$ and
$\tau\mupn$  represents the stress-energy of tensor matter in
the bulk. Although we will consider an empty bulk in what follows,
we still wrote down its contribution to understand the way bulk
matters couple to gravity.
To concentrate on our specific model, we consider
the bulk action \eqref{action}. This corresponds to the previous
action $S^{(\xi, \beta)}$, where now the parameters $\xi$ and $\beta$
have been fixed to $\xi=-1$ and $\beta=2$. The overall coefficient is
therefore $\xi\(1-\beta\)=+1$. This corresponds to a double flip of
sign such that the bulk equation of motion is indeed {\bf
  unchanged} and has the conventional sign:
\be
\left[ \partial_y^2 +\boxdot -\left(\frac{9}{4}+\frac{15}{4\sinh^2
    Hy} \right)H^2 \right] h\mupn=-\frac{\kf}{2}\tau \mupn.
\ee
Gravity couples the right way to matter in the bulk, even though the
gravity part of the bulk action comes in with an unconventional
sign, this is due to the presence of the GB terms that dominate over
the bulk EH term.
The bulk geometry is therefore stable against the introduction
of matter in the bulk, at least at the linearized level, unlike the models presented in Refs.
\cite{Deser:1989jm,deRham:2006hs} which were unstable due to the
presence of relatively large GB terms. In this new model, the
overall unconventional sign makes the theory stable in presence of
large GB terms.
In what follows, we set $\tau \mn=0$. 

It has been pointed out in Ref. \cite{Hebecker:2001nv} that the presence of hot matter on
the brane in the case of an unconventional GB AdS solution leads to a
naked singularity in the bulk. However our scenario is slightly
different to that considered in Ref. \cite{Hebecker:2001nv},
(in particular the bulk scalar curvature does not
dominate over the GB term). It is therefore unclear whether the
same result is valid here. 

Following the same procedure as section \ref{tensors1}, we may work
in the GN gauge \eqref{h GN y} to derive the boundary conditions. Here
again, the boundary conditions should be modified by an overall factor
$\xi(1-\beta)$ which is simply $+1$ for our specific model.
Working back in terms of the RS gauge, we therefore get
\ba
 \left[\partial_y +\frac{3H}{2}\coth H \by +
 \(\frac{\alpha \kf}{2\kq}-2\l^2 H \coth H \by \) \boxdot
 \right] h\mupn&=&-\kf\(\delta T\mupn-\frac 1 3\,  \delta T \delta\mupn\) \label{bdy GB}\\
 &&\hspace{-40pt}+ \left[2+\frac 1 H \coth H \by
   \(\frac{\alpha\kf}{\kq}-4\l^2 H \coth H \by\)\boxdot\right]
 \hat{F}\mupn \delta
 y .\notag
\ea
The trace of this equation imposes the equation of motion for the
brane displacement:
\ba
\(2+\frac 1 H \coth H \by \(\frac{\alpha\kf}{\kq}-4\l^2 H \coth H
\by\)\boxdot
\)\left[\Box+4H^2\right]\delta y=-\frac{\kf}{3}\,  \delta T.
\ea
As mentioned in section \eqref{tensors1}, in the absence of tensor
matter sources on the brane, $\delta y$ is
such that $\left[\Box+4H^2\right]\delta y=0$. Thus the tensor $\hat{F}\mn \delta
y$ satisfies the same equation of motion \eqref{eq d y} as in the previous
model. Since $\by>0$ in this model, its perturbed boundary equation
is very similar to the one of the RS$_+$ model which is stable. We
therefore expect to recover the same features in this model.

\subsubsection{Tensors}
We focus for now on tensor perturbations in the vacuum ({\it i.e.} $\delta T\mn=0$). To start with, we do
not consider the scalar contribution in the second line of \eqref{bdy GB}.
Performing the decomposition $h_{\mu\nu}=\sum_m u_m(y) \chi_{\mu\nu}^{(m)}$ we obtain
\ba
&&\left[ \partial_y^2 +m^2 -\left(\frac{9}{4}+\frac{15}{4\sinh^2 H y}
  \right)H^2 \right] u_m(y)=0, \label{u''}\\
&&\left[\partial_y+\frac{3H}{2} \coth
  H\bar{y}+\left(\frac{\alpha\kf}{2\kq}
  -2 \l^2 H \coth H \bar y
  \right)m^2 \right] u_m(\bar{y})=0. \label{bdy u}
\ea
Again it is straightforward to see that the graviton zero mode is massless $m=0$ with profile
\be
u_0(y)=A(\sinh H y)^{-3/2}.
\ee
The remaining modes are massive with $m^2>9H^2/4$. A more detailed
  analysis of the solutions of these equations can be found in
  appendix \ref{eigenvalues}.

\subsubsection{Brane-bending mode}

The fact that we have found the graviton zero mode to be massless
guarantees that there are no zero helicity graviton modes that could play the r\^ole of ghosts.
Furthermore as we now explain, there is no brane-bending mode either
and hence the modes which are dangerous in the DGP context, decouple
in the present scenario. 
Following the discussion of section \ref{tensors1}, in the vacuum the
brane displacement satisfies the same equation of
motion \eqref{eq d y} as in the previous section. It can therefore
be assimilated to a tensor mode with $m^2=2H^2$ and can thus in principle
source a similar mode in the bulk. However, because we are interested
in the same range as in the RS$_+$ model, the normalizable mode with
$m^2=2H^2$ behaves identically to the brane-bending mode, {\it i.e.}
$u_2(y)\sim \cosh H y / \(\sinh H y\)^{3/2}$. Applying the boundary
condition
simple enforces the combination of the tensor mode and the brane displacement to
vanish. Hence this mode does not become physical.

Another way to argue for this is following appendix \ref{eigenvalues},
since the normalizable mode with $m^2=2H^2$ is positive
definite everywhere, it cannot be orthogonal to the zero mode which is
similarly positive definite. But since orthogonality of
different eigenmodes with different eigenvalues is guaranteed, this
mode cannot exist.

\subsection{Response to brane matter: Modifications of Newton's law at
subhorizon scales}
\label{mod gravity}

In what follows we study the response to a matter source $\delta T\mn$ on the
brane.
At very small scales compared to the Hubble radius $r\ll H^{-1}$, the
geometry does not feel the dS expansion and appears
Minkowski-like. Neglecting $H$ at these scales, one can solve the mode
equations exactly
\ba
h\mn=\frac{\sqrt{y/\l}K_2(\sqrt{-\Box}y)}{K_1(\sqrt{-\Box}\l)+2\sqrt{-\Box}\l
\(\frac{\gamma}{6}-1\)
K_2(\sqrt{-\Box \l})}\frac{\kf}{\sqrt{-\Box}}\, \Sigma\mn,
\ea
where $K_n$ are the Bessel functions and the transverse and traceless
part of the stress-energy is
$\Sigma\mn=\delta T\mn-\frac 1 3\,
\delta T\gamma\mn+\frac{1}{3 \Box} \delta T_{;\, \mu \nu}$.
The induced metric perturbation on the brane $\bar{h}\mn$ in
de Donder gauge is therefore
\ba
\hspace{-25pt}\bar{h}\mn=
-\frac{2 \kq}{\Box}
\(\delta T\mn-\frac 1 2 \, \delta T \gamma\mn\)
+
\frac{K_0(\sqrt{-\Box}\l)}{
\frac{\kf}{\l \kq}
K_1(\sqrt{-\Box}\l)+\sqrt{-\Box}\l
\(\frac{\gamma}{3}-2\)
K_0(\sqrt{-\Box}\l)}
\frac{\l \kq}{
\sqrt{-\Box}}\Sigma\mn .
\ea

\begin{itemize}
\item {$\lambda \ll \l \ll H^{-1}$}
\end{itemize}
At short-distances compared to $\l$, {\it i.e.} for $\sqrt{-\Box}\l\rightarrow
\infty$, the induced perturbations on the brane follow a behavior
very similar to the usual four-dimensional one:
\ba
\bar{h}\mn\rightarrow -\frac{2\kq}{\Box}\(\delta T\mn-\frac 1 2 \, \delta T \gamma\mn
+\frac{1}{\(\frac{2}{3}\gamma-4\)} \Sigma\mn\).
\ea
In fact we can write this equation in the conventional form
$\delta G\mn=\kq  \,\delta T^{\text{eff}}\mn$,
where the effective
stress-energy is defined by
\ba
\delta T^{\text{eff}}\mn=\delta T\mn+\frac{1}{\( \frac{2}{3}\gamma-4\)}
\(\delta T\mn-\frac{1}{3}\, \delta T\gamma\mn+\frac{1}{3\Box}\delta T_{;\, \mu\nu}\), \label{smallscales}
\ea
which relates in a non-local way to the stress-energy $\delta T\mn$.
If $\gamma$ is large enough, (as will be the case when we put numbers in in section \ref{implications})
no modification to gravity will therefore be observable at scales smaller that
$\l$. The fact that we recover four-dimensional gravity at small
distance in presence of a EH term in the brane was pointed out in Ref.
\cite{Collins:2000su}, (note that GB terms in the bulk have a similar
effect).

For subhorizon wavelengths much longer that $\l$ {\it i.e.} $\l^2\Box\rightarrow
0$, we have
$\delta G\mn=\kq  \,\delta T^{\text{eff}}\mn$, where now
\be
\delta T^{\text{eff}}\mn=\delta T\mn+\frac{3}{4\( \gamma-\frac{9}{2}\)}\(\Gamma_0+\log
  \frac{\sqrt{-\Box}\l}{2}\)\, \l^2 \Box \Sigma\mn +\mathcal{O}(\l^4\Box^2 \delta T\mn),
\ee
where $\Gamma_0$ is the Euler's constant, ($\Gamma_0\simeq0.577$).
In this regime, the corrections to four-dimensional gravity are suppressed by
$\l^2\Box $ and will additionally be suppressed if $\gamma$ is
sufficiently large.

The fact that we recover four-dimensional gravity on the brane at all subhorizon scales if $\gamma$
is sufficiently large follows from the fact that $\gamma$ measures the ratio of the brane
EH term to the bulk EH term. When $\gamma$ is large the brane term dominates and we recover four-dimensional
 gravity with corrections suppressed by $1/\gamma$.
 In appendix \ref{ppwaves}, we show that this remains true at the
 nonlinear level.

\section{Cosmological Implications}

\label{implications}

\subsection{Observed cosmic acceleration}

Our main motivation is to use the emergent acceleration presented in
section  \ref{selfGB}
to describe the current cosmic acceleration. Thus
the effective cosmological constant \eqref{cc2} should be of order
$\Lambda_{\text{eff}}\sim \bar{r}/ \kq \l^2 \sim 1/\kq l_H^2$
where $l_H$ is the
current Hubble scale (size of our Universe): $l_H\sim
10^{43}$GeV$^{-1}\sim 10^{10}$ light-years. The AdS curvature should
therefore be $\l\sim l_H/\sqrt{\bar{r}}$. In the regime where
$\gamma\gg 1$, $\bar r  \sim 1/\gamma$.

Our main constraint is the fact the higher orders $\rho_{\text{higher}}$ in the
Friedmann equation \eqref{H2} are negligible $\rho_{\text{higher}}\ll
  \rho$ {\it i.e.} $\rho\ll \(\gamma-\frac 9 2\)/ \l \kf\sim 1/\l^2 \kq$,
or equivalently,
that $\l H \ll 1$.
The initial theory has three free parameters $\kf$, $\l$ and
$\alpha$. As seen in section \ref{selfGB}, $\alpha$ is fixed such that
we recover the conventional coupling to matter in the
Friedmann equation. Obtaining the correct
late time cosmological constant will then fix the combination $\l \kf$.
Furthermore, if we demand that the
low-energy limit $\l H\ll 1$ or
$\rho\ll 1/\l^2 \kq$ is valid all the way back to nucleosynthesis
($\rho_n \sim \(10^{-3}\text{GeV}\)^4$),
then we have the additional constraint
\be
\l \ll \l_c= H_n^{-1} \sim 1/\kappa_4 \sqrt{\rho_n} \sim 10^{25} \, \text{GeV}^{-1}.
\ee
This leads to the fine-tuning of the parameters $1/\bar r\sim \gamma \gg l_H^2/\l_c^2 \sim
10^{36}$. In this limit we therefore have $\alpha=1+9/2 \gamma \sim
1$, and the constraint $\gamma\gg \gamma_c$ (necessary
for the existence of an accelerating solution) is easily satisfied. Since $\gamma\gg1$, the
five-dimensional corrections to four-dimensional gravity are therefore strongly suppressed.

This model can give rise to an effective dark energy
component which can be observationally acceptable.
However we see that taking seriously the tuning of this model,
the AdS length scale could be unusually large: $\l < \l_c= 10^3$km,
{\it i.e.} of a scale at which numerous
observational tests can be performed. One might be
worried of the modification of gravity at scales of the same order of
magnitude or below. But as pointed out in \eqref{smallscales}, if
$\gamma$ is large enough (and in our model it could be at least
$10^{36}$),
the modifications of gravity at subhorizon scales
will be unobservable.

\subsection{Five-dimensional Planck scale}

The low-energy constraints leads to a
five-dimensional gravitational coupling constant $\kf \gg \kappa_c^2$,
where $\kappa_c^2=l_H^2 \kq /\alpha \l_c \sim \(10^{-8}\text{GeV}\)^{-3}$.
This corresponds to a remarkably low five-dimensional Planck mass of about $10 \, \text{eV}$.
The fact that the five-dimensional Planck scale is so small, suggests that
we would be able to easily produce small black holes in current laboratory experiments,
something which is clearly ruled out. However we argue that this is not necessarily the case.
Consider the collision of two high energy particles on the brane, the total stress-energy of
the brane is not given just by the stress-energy of matter, but by the combination
\be
\hat{T}\mn=T\mn-\frac{1}{\kq} G\mn .
\ee
As is clear from the arguments given so far, (e.g. equation \eqref{smallscales})
these two contributions cancel out at leading order and give rise to
\be
\hat{T}\mn \sim \frac{1}{\gamma} T\mn \sim \frac{\l \kq}{\kf}
T\mn .
\ee
Thus for large $\gamma$ the effective stress-energy on the brane is a negligible fraction
of the real matter stress-energy. Indeed it is suppressed by the same factor of $1/\kf$ as the bulk action.
Thus the system behaves like the RS model but with an effective five-dimensional gravitational coupling
$\tilde{\kappa}_5^2=\kq \l$ which corresponds to a five-dimensional
Planck mass of an acceptable $\tilde{M}_5=10^{4}-10^5 \ \text{GeV}$.
We expect then that the production rate of black holes will be
determined by this scale, which is sufficiently
high to have evaded current accelerator experiments. In appendix C we
give some nonlinear solutions describing high energy particles which
exhibit this effect.
If additional dimensions are compactified on very small scales the fundamental
higher-dimensional Planck scale could again be much larger.

\subsection{Finite maximal energy density}

To finish, we remark briefly, on the ``maximal energy density''
feature generic to RS-GB models, as pointed out in
Refs. \cite{Brown:2005ug}. Considering the Friedmann equation
\eqref{FriedGB}, there exist a maximal {\it finite}
energy density
\ba
\rho_\text{max}=\frac{\gamma^3}{54 \l \kf}+\mathcal{O}(\frac{\gamma}{\l\kf}),
\ea
for which the Hubble parameter is $H^2\sim\gamma^2/\l^2$. If this
model was trusted all the way back to the Big Bang time, the Big Bang
temperature, scale factor and Hubble constant would be finite. The
solution is nevertheless singular since $\dot{H}$ diverges, this is a
very mild unusual spacelike singularity. Although this maximal value
would be huge, ($\rho_\text{max}\sim
(10^{15}\text{GeV})^4$ with the previous fine-tuning), it is nevertheless
comparable to the energy scale typically taken at the beginning of
inflation ($\rho_{\text{beg}} \sim (10^{16}\text{GeV})^4$, see Ref. \cite{LL}) which suggests that the
inflation should start at much lower energy in this model.  The end of
inflation is usually taken to be at around
$10^{13\pm3}\text{GeV}$, so this model could strongly affect the
behavior of some inflationary models.
The study of this feature is however beyond the scope of this paper and we suggest
Ref. \cite{Brown:2005ug} for further details on this interesting issue.

\section{Perspectives}
\label{secsum}

We have presented a model of gravity which accelerates in the absence of any cosmological constant or potential which can
provide an equivalent description of most of cosmic history without
the need for the introduction of a cosmological constant. There are
many questions that this model raises. Firstly, the formal: Is the
bulk theory stable beyond linear order? If not what do the
instabilities imply? Are there other higher derivative actions with
similar properties? In particular, can we extend it to models where
the bulk action takes the conventional sign but with higher order
corrections?
It has recently been demonstrated that higher derivative corrections
to gravity alone, to all orders in the $\alpha'$ expansion, give rise to AdS
solutions in string theory, even in the absence of a cosmological
constant term induced from the matter sector, see Ref.
\cite{Friess:2005be}. Our model could be viewed as a toy version of
these.
Can such an action be derived from a UV complete
theory? Secondly, the phenomenological: How do the high energy
modifications of gravity affect inflation, structure growth? How well
can we constrain the free parameter $\l$? And finally how can the
different cosmological tests performed on the DGP model (see for
instance Refs. \cite{Maartens:2006yt}) be extended to
this scenario?


\section*{Acknowledgements}
We would like to thank C.~Charmousis, J.~Cline, A.~Davis, C.~Fang, R.~Gregory, N.~Kaloper,
S.~Kanno, J.~Khoury, T.~Padilla, T.~Shiromizu,
P.~Steinhardt, A.~Upadhye and T.~Wiseman for useful discussions.
CdR is funded by a grant from the Swiss National Science
Foundation. AJT is supported in part by US Department of Energy Grant
DE-FG02-91ER40671.

\appendix
\setcounter{equation}{0}
 \renewcommand{\theequation}{\thesection.\arabic{equation}}

\section{Eigenvalues and eigenfunctions for the new model}

\label{eigenvalues}

Let us now analyse the normalizable tensor perturbations for the new model.
To simplify the equations of motion and boundary conditions we define
$\mu=m/H$, $x=H y$. The brane position is therefore at $\sinh
\bar{x}=\l H$. The the bulk equation then
becomes
\be
\label{eom1}
\left[ \partial_x^2+\mu^2-\(\frac{9}{4}+\frac{15}{4  \sinh^2x} \)
\right]u_m(x)=0,
\ee
and the boundary condition is
\be
\label{bound2}
\left[ \partial_x+\frac{3}{2} \coth x +\frac{\mu^2}{3}  \coth x
  (1-2\sinh^2 x)\right]_{x=\bar{x}}
\hspace{-10pt}u_m(\bar{x})=0.
\ee
Here we have made explicit use of the background equations of motion
when the brane is empty, {\it i.e.} Eq. \eqref{FriedGB}.
The inner product is defined
as
\be
(f,g)=\int_{\bar{x}}^{\infty} \d x f(x) g(x) + \frac{1}{3} \coth
\bar{x} (1-2\sinh^2 \bar{x})  f(\bar{x})g(\bar{x}),
\ee
and one can demonstrate that non-degenerate eigenfunctions will be
orthogonal with respect to this definition.
To ensure that the norm is positive definite we require that
$\sinh{\bar{x}} \le 1/\sqrt{2}$ which is the case in our model.
For $\mu^2>9/4$ there are a continuum of normalizable modes which
behave like a superposition of plane waves at the horizon. In the
range $0 \le \mu^2<9/4$ normalizability requires that the modes
fall of as $\exp(-\bar{\mu} x)$ as $x \rightarrow \infty$ where
$\bar{\mu}=\sqrt{9/4-\mu^2}$. The solution Eq. \eqref{eom1} with this
property is
\ba
\label{modes1}
u_\mu(x)=A_\mu \sqrt{\sinh x}  \ Q^{(2)}_{\nu}\(\cosh x\),
\ea
where $Q_{\nu}^{(n)}$ is the associated Legendre function and
$\nu=-\frac{1}{2}+\sqrt{\frac 9
  4-\mu^2}$.
The existence of a mode in
the region depends on whether this function satisfies the boundary
condition (\ref{bound2}). It is straightforward to show that for $\mu^2=0$ the solution is
\be
u_0(x)=\frac{A_0}{(\sinh x)^{3/2}}
\ee
and that this is normalizable and consistent with the boundary
condition. Now for any normalizable function for $\mu^2 \le 9/4$ we
have $\partial_x^2u/u \ge 0$ and as $x \rightarrow \infty $, $\ln u
\rightarrow -\sqrt{9/4-m^2}$. Assuming $u>0$ the gradient $\partial_x
u$ increases with increasing $x$, which in turn implies $\partial_x
u<0$ for all $x$ and hence the functions \ref{modes1} have no zeros,
{\it i.e.} there is no real solution to $u(x)=0$.
 Since the definition of the inner
product is positive definite, it would be impossible for such a mode
to be orthogonal to the zero mode, and hence the the only normalizable
mode consistent with the boundary conditions in the entire range
$\mu^2 \le 9/4$ is the zero mode. This result is crucial because it
implies both that there are no ghost states arising from the dangerous
regime $0<\mu^2<2$ and no unstable modes with $\mu^2<0$.

\section{Gravity in the RS$_-$ model}
\label{app neg}
In this section, we evaluate the mass of the zero mode in the
RS$_{-}$ model
of section \ref{tensors1}. Using the same notation as the previous
appendix, the bulk equation of motion is the same as \eqref{eom1}, and
for $\mu^2<9/4$, the solution of the bulk
equation which is normalizable in the range $\bar{x}<x<0$ is
\ba
u_\mu(x)=A_\mu \sqrt{\sinh x}  P^{(2)}_{\nu}\(\cosh x\),\label{all
m}
\ea
where $P_{\nu}^{(n)}$ is the other associated Legendre function.
This solution  is valid if $\mu^2 \ne 0$ and $\mu^2 \ne 2$.
For $\mu^2=0$, the
normalizable mode in the region $\bar{x}<x<0$ is
\ba
u_0(x)=A_0\, \frac{\cosh^2 \! x-3 \cosh x+2}{\(\sinh x\)^{3/2}}, \label{m=0}
\ea
 and for $\mu = 2$,
the normalizable mode is
\ba
u_2=A_2 \, \frac{\(\cosh x -1\)^2}{\(\sinh x\)^{3/2}}. \label{m=2}
\ea
If we consider the brane to be empty, $\rho=0$, the Hubble parameter
is given in \eqref{H -} and the boundary condition is therefore
\ba
\left[\partial_x+\frac{3}{2}\coth \bar{x}+\mu^2 \sqrt{1-\frac 2 3
    \gamma}\right]u_m(\bar{x})=0,\label{bdy -}
\ea
where $\sinh \bar{x}=-\frac{2}{\gamma}\sqrt{1-\frac{2}{3}\gamma}$. We
can easily check that both \eqref{m=0} and \eqref{m=2} never satisfy this boundary
condition no matter what $\gamma$ is. However, when $\mu^2=2$, the
mode can couple to the brane-bending mode. Effectively , the presence of a
brane-bending mode with $\mu^2=2$, appears as a source term in the
boundary condition \eqref{bdy -}. This possibility is explored in section
\ref{tensors1} and is shown to be the source of an instability on this
brane.

Using the more general solution
\eqref{all m} on the other hand, we see that the boundary condition
\eqref{bdy -} is satisfied for $\mu^2=\mu^2(\gamma)>2$, where its exact value
depends on $\gamma$. In this model, despite the fact that the zero
mode is massive, it does not lie in the forbidden region $0<\mu^2<2$,
provided no matter is introduced on the brane. For $H$ greater than
the vacuum value, \eqref{H -}, the value of $\mu^2(\gamma)$
decreases and crosses $2$ when $H$ is greater than a critical value
{\it i.e.} when enough matter is introduced on the brane. The model
becomes then unstable. Although this self-accelerating branch appears
stable under a critical value of the energy density on the brane, we
can show in what follows that gravity cannot be localized on the brane.

At subhorizon scales, the Hubble parameter is negligible, and the equations
of motion become
\ba
&&\left[\partial_y^2+\Box-\frac{15}{4 y^2}\right]\, h\mupn=0\\
&&\left[\partial_y+\frac{3}{2 \bar{y}}+\frac{\alpha \kf}{2\kq}
  \Box \right]\, h\mupn(\bar{y})=-\kf T\mupn,
\ea
where $\bar{y}=\l<0$ and $T\mn$ is the stress-energy tensor
introduced on the brane, which is assumed to be traceless for simplicity.
The normalizable mode function in the range $\l<y<0$ is $h\sim
\sqrt{y} I_2(\sqrt{-\Box} y)$ where $I_n$ is the Bessel I function.
On the brane, the mode functions are therefore:
\ba
h\mn(\bar y)=\kf\frac{I_2(\al\sqrt{-\Box})}{\sqrt{-\Box} I_1(\al\sqrt{-\Box})-\frac{\alpha \kf}{2
\kq}\Box I_2 (\al \sqrt{-\Box})}T\mn,
\ea
so that the perturbed metric on the brane is
\ba
h\mn=\kf \left[\frac{\al}{4}
-\frac{1}{96}\(\al +3\,\frac{\alpha \kf}{\kq}\)\l^2\Box+\cdots
\right]T\mn(x^\mu).
\ea
Thus the induced perturbations on the brane don't satisfy the ordinary
four-dimensional equation, and gravity will not appear localized on
this negative tension brane. This can be understood easily by the fact
that the warping factor increases exponentially away from the brane
and gravity will tend to be localized at the boundary of AdS $y=0$, {\it i.e.}
infinitively far away from the brane, unlike the positive tension
brane case.

Writing $\frac{\alpha \kf}{\kq} \sim
r_0$, the cross over-scale, one however recovers the standard DGP
result in the limit $\al \rightarrow \infty$
\ba
h\mn(\bar y)\rightarrow \frac{\kf}{\sqrt{-\Box}-r_0 \, \Box} T\mn,
\ea
such that at very small scales $k\gg 1/r_0$, one recovers the
four-dimensional standard behavior $h\mn \sim -\frac{\kq}{r_0 \Box} T\mn$,
whereas at large cosmological scales, gravity becomes five-dimensional.

\setcounter{equation}{0}

\section{Nonlinear gravitational waves}

\label{ppwaves}

An ultra-relativistic particle may be described by a brane stress-energy of the form $T_{uu}=E \delta^2(x)\delta(u)$, where $E$ is its
energy. For $\l \ll H^{-1}$ we can find an exact nonlinear solution of
the bulk equations in the form of a warped pp-wave
\be
\d s^2=a^2(y)\(\d y^2+2\d u\d v+a^{-3/2}(y)f(y,u,x_2,x_3)\d u^2+\d x_2^2+\d x_3^2\),
\ee
where $u=(x_1-t)/\sqrt{2}$, $v=(x_1+t)/\sqrt{2}$ and
$a(y)=\l/y$. This is a solution provided both the bulk equation and
the boundary condition
\ba
&&\left[\partial_y^2+\Box-\frac{15}{4 y^2}\right] f(y,u,x_2,x_3) =0\\
&&\left[\partial_y   +\frac{3}{2 \l} +\(\frac{\alpha \kf}{2\kq}-2
\l\)
\Box \right] f(\bar{y},u,x_2,x_3) =-\kf\, T_{uu}
\ea
are satisfied.
Here $\Box=\partial_{x_2}^2+\partial_{x_3}^2$ and the brane is
located at $y\equiv\bar{y}=\l$. In fact this is identical to the special
case of linearized perturbations considered earlier, except the solution is now valid
{\it non-linearly} for $f$ of arbitrarily large
amplitude. The general solution for arbitrary source $T_{uu}$ is
\ba
f=-\frac{3 \kf}{\l \(\gamma -3 \)}\left[
1+\frac{\l \sqrt{-\Box}}{2}
\frac{K_0(\l \sqrt{-\Box})}{\(\frac{2\gamma}{3}-1\)K_1(\l
 \sqrt{-\Box})
+\(\frac{\gamma}{3}-2\) \l  \sqrt{-\Box}K_0(\l \sqrt{-\Box})}
\right]\frac{1}{\Box}T_{uu} .
\ea
Similarly as in section \ref{mod gravity}, at very small scales, $\l^2 \Box\gg 1$, this
simplifies to:
\ba
f_{(r\ll \l)}\approx-\frac{3\(\gamma-\frac 9 2\)}{\(\gamma-\frac 3
 2\)\(\gamma-6\)} \frac{\kf}{\l \, \Box} T_{uu},
\ea
whereas at scales much larger than $\l$, $f$ simplifies to
\ba
f_{(r\gg \l)}\approx-\frac{3 \kf}{\l \(\gamma -3 \)}
\left[\frac{1}{\Box}
+\frac{\l^2}{2\(\frac{2\gamma}{3}-1\)}
 \(\Gamma_0+\log \frac{\l \sqrt{-\Box}}{2}
\)+\cdots
\right]T_{uu}.
\ea
The solution corresponding to $T_{uu}=E
\, \delta^2(x)\delta(u)$ is the five-dimensional version of the Aichelberg-Sexl metric
\cite{Sexl} generalized to the present model. Similar shock-wave
solutions in the DGP context have been considered in
\cite{Kaloper:2005az,Kaloper:2005wa}. In particular at small 
scales, $f$ is
\ba
f_{(r\ll \l)}&\approx& \frac{3\(\gamma-\frac 9 2\)}{\(\gamma-\frac 3
 2\)\(\gamma-6\)} \frac{\kf}{ \pi \l} \, E \delta(u)\, \log \frac{r}{\l} \notag\\
&\approx& \frac{\(\gamma-\frac 9 2\)^2}{\(\gamma-\frac 3
 2\)\(\gamma-6\)} \frac{2 \kq}{\pi} \, E \delta(u)\, \log
 \frac{r}{\l},
\ea
where $r=\sqrt{x_2^2+x_3^2}$. At small scales, we therefore recover the standard four-dimensional result, up
to a factor which tends to $1$ when $\gamma \gg 1$.

At larger scales, when $H^{-1}\gg r\gg \l$, the corrections to the
standard result are suppressed by a factor of $\frac{\l}{ r \gamma} \ll 1$:
\be
f_{(r\gg \l)}\approx \frac{\kq \(\gamma-\frac 9 2\)}{ \pi \(\gamma -3 \)}
\left[ \log \frac{r}{\l}+\frac{\Gamma_0}{2\(\frac{2\gamma}{3}-1\)}\frac{\l^2}{r^2}\Lambda
\right] E\, \delta(u) ,
\ee
where $\Lambda$ is the cut-off scale. Typically in this
approximation, we can take it to be $\Lambda \sim r/\l$, so that the
correction term is suppressed by both a factor of $1/\gamma$ and a
factor of $\l/r$.

As long as $\gamma \gg 1$ this solution is essentially equivalent to the conventional
four-dimensional Aichelberg-Sexl metric on the brane. Since this stress-energy
source is traceless, the position of the brane remains unperturbed by
the gravitational wave.
This solution differs significantly from that say in the RS model,
where for distances much less than $\l$ we recover the five-dimensional version of
the Aichelberg-Sexl metric. This confirms our general argument that
the production rate of black holes from high energy collisions is
suppressed.

\end{document}